\documentclass[journal=jpccck,manuscript=article]{achemso}

\pdfoutput=1

\usepackage{graphicx}
\usepackage{amsfonts}

\author{Alexei V. Khomenko}
\email{khom@mss.sumdu.edu.ua}
\author{Nikolay V. Prodanov}
\affiliation[Sumy State University]
{Department of Modeling of Complex Systems, Sumy State University, 2, Rimskii-Korsakov St. 40007, Sumy, Ukraine}

\title[Study of friction of Ag and Ni nanoparticles]
{Study of friction of Ag and Ni nanoparticles: \\ an atomistic approach}

\begin{document}
%%%%%%%%%%%%%%%%%%%%%%%%%%%%%%%%%%%%%%%%%%%%%%%%%%%%%%%%%%%%%%%%%%%%%
%% The manuscript does not need to include \maketitle, which is
%% executed automatically.  The document should begin with an
%% abstract, if appropriate.  If one is given and should not be, the
%% contents will be gobbled.
%%%%%%%%%%%%%%%%%%%%%%%%%%%%%%%%%%%%%%%%%%%%%%%%%%%%%%%%%%%%%%%%%%%%%
\begin{abstract}
  Manipulation of metal nanoparticles using atomic force microscope is a promising new technique for probing tribological properties at the nanoscale. In spite of some advancements in experimental investigations, there is no unambiguous theoretical treatment of processes accompanying the movement of metallic nanoislands adsorbed on a flat surface and additional research is required. In this paper, we describe computer experiments based on classical molecular dynamics in which the behavior of silver and nickel nanoparticles interacting with a graphene sheet and sheared with constant force is studied. Frictional force acting on the nanoislands is measured as a function of their size. It is shown that its average value grows approximately linearly with contact area, and slopes of linear fits are close to the experimentally observable ones. The dependence of the friction force value and of the shape of the measured friction curves on the type of metal atom is revealed and its possible reasons originating from atomistic background are discussed.

\end{abstract}

\section{Introduction}
\label{intro}

Due to modern trends towards the intensive miniaturization of various devices with moving constituents friction and wear at the nanoscale is more and more often dealt with in technological applications and in many instances these phenomena are the main obstacle to achieving their reliable functioning~\cite{Gnecco2009,Pantazi2008,Bhushan2008,Socoliuc2006}. Although traditional experimental techniques, such as friction force microscopy (FFM) and surface force apparatus (SFA) have significantly enhanced the understanding of atomistic origins of friction, these methods have some deficiencies. In particular, FFM lacks the possibility of direct and independent measurement of the true contact area of the sliding interface, and both mentioned techniques are not able to probe tribological properties of contacts with contact area in the range from about several hundreds to some hundred thousands of nm$^2$~\cite{Gnecco2007}. A promising new approach capable of solving mentioned difficulties is the study of frictional properties of adsorbed nanoparticles by moving them with the tip of an atomic force microscope (AFM)~\cite{Gnecco2007}. A variety of experimental works concerned with the manipulation of nanoislands can be found in literature~\cite{Gnecco2007,Junno1995,Resch1998,*Baur1998,Sitti2000}. However, most of them are targeted only at dislocation of the nanoparticles and do not investigate tribological processes. Quantified frictional properties of antimony nanoparticles grown on highly oriented pyrolytic graphite (HOPG) and pushed with the tip of an AFM have been described recently~\cite{Gnecco2007,Dietzel2008,Ritter2005}. These experiments indicate the linear dependence of friction force on the contact area and the main feature found is that islands with area less than about $10^4$~nm$^2$ are much easier to move than the ones with larger contact areas. Additionally, in the last work~\cite{Dietzel2008} the authors revealed the so-called ``frictional duality'', where some particles with the areas of more than $10^4$~nm$^2$ assume a state of frictionless sliding while the others show finite friction. However, there is no clear explanation of the observed behavior. In earlier works~\cite{Gnecco2007,Ritter2005} lower friction of the smaller nanoparticles is mainly attributed to their compact amorphous structure and incommensurability with the substrate in contrast to larger nanoislands, which are often ramified and would not move as rigid entities thus providing new route for energy dissipation. While in Ref.~\cite{Dietzel2008} the coexistence of the two frictional states is ascribed to the presence of contamination molecules and structure of the nanoparticles is not considered as a crucial factor defining the observed behavior.

Mentioned ambiguity indicates the need of additional, in particular, theoretical investigations. Existing analytical or seminumerical models~\cite{Gnecco2007,Ritter2005,Aruliah2005} can provide some estimates of the experimentally observed quantities. But they are often based on a large amount of assumptions and thus may not be able to yield clear picture of the multiple-factor tasks due to neglecting some of the contributions which can play appreciable role. Large-scale computer simulations using classical molecular dynamics (MD) is an alternative approach capable of elucidation of the detailed atomistic behavior of the system and may help to reveal some subtle effects inaccessible for the analytical techniques. In literature, however, MD simulations pertaining to the movement of metallic nanoislands on a graphite surface are mainly concerned with their diffusion and thus nanoclusters consisting of up to several hundreds of atoms are investigated~\cite{Luedtke1999,Lewis2000,Yoon2003}. Such small systems are not appropriate for the study of tribological properties and nanoparticles of several orders of magnitude larger should be inspected.

The absence of computational investigations of friction of relatively large metallic nanoparticles on a graphitic surface provides the impetus for the current study. In this work we report MD simulations exploring friction of Ag and Ni nanoparticles with up to 30,000 atoms on a graphene sheet. Such sizes of the clusters are smaller than in the tribological experiments described above. Nevertheless, they are suitable for the study of friction (they do not diffuse at room temperature) and understanding of their behavior may gain some valuable insights into tribological processes. The choice of the metals is caused by the absence of the reliable interaction potential for antimony. Properties of Ag (lattice constant, mass, density) are the closest to the ones of Sb among metals with face-centered cubic (FCC) lattice~\cite{MetalsHandbook} available from the database of the used potential form. Ni is chosen as a trial material for the exploring the influence of type of metal on the system behavior. Additionally, novel experimental techniques of synthesis of Ag~\cite{Jeon2010} and Ni~\cite{Geissler2010} nanoparticles with sizes comparable with our model are rapidly developed making these materials possible candidates for the future manipulation experiments allowing direct comparison with the results of the simulations. The use of a single graphene sheet in spite of graphite consisting of multiple graphene layers can be considered as the first approximation towards approaching the experimental conditions. Deeper understanding of friction and wear of graphene interacting with various materials may also be valuable, as these processes are often related to graphene production~\cite{Khomenko2010,Prodanov2010,Sen2010}. Moreover, in literature there is a growing interest directed at the interactions of a graphene sheet with different nanoobjects as they can change electronic properties and structure of this material, which may have an implication for future nanodevices~\cite{Shibuta2006,Neek2009,Crespi2009}. The main objectives of our study are to define the influence of the size, structure of a nanoparticle and of the type of metal on the friction force and to elucidate atomistic processes occurring during the shear of the nanoislands in the context of our model.

\section{Model}
\label{model}

\begin{figure}[htb]
\centerline{\includegraphics[width=0.51\textwidth]{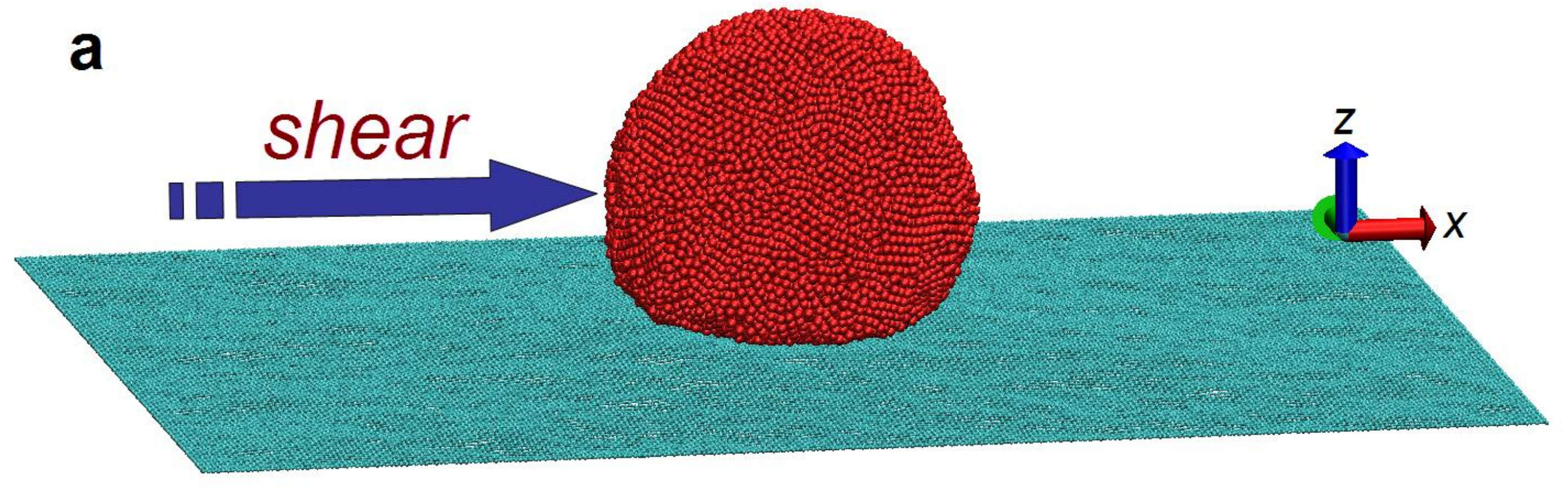}
\includegraphics[width=0.51\textwidth]{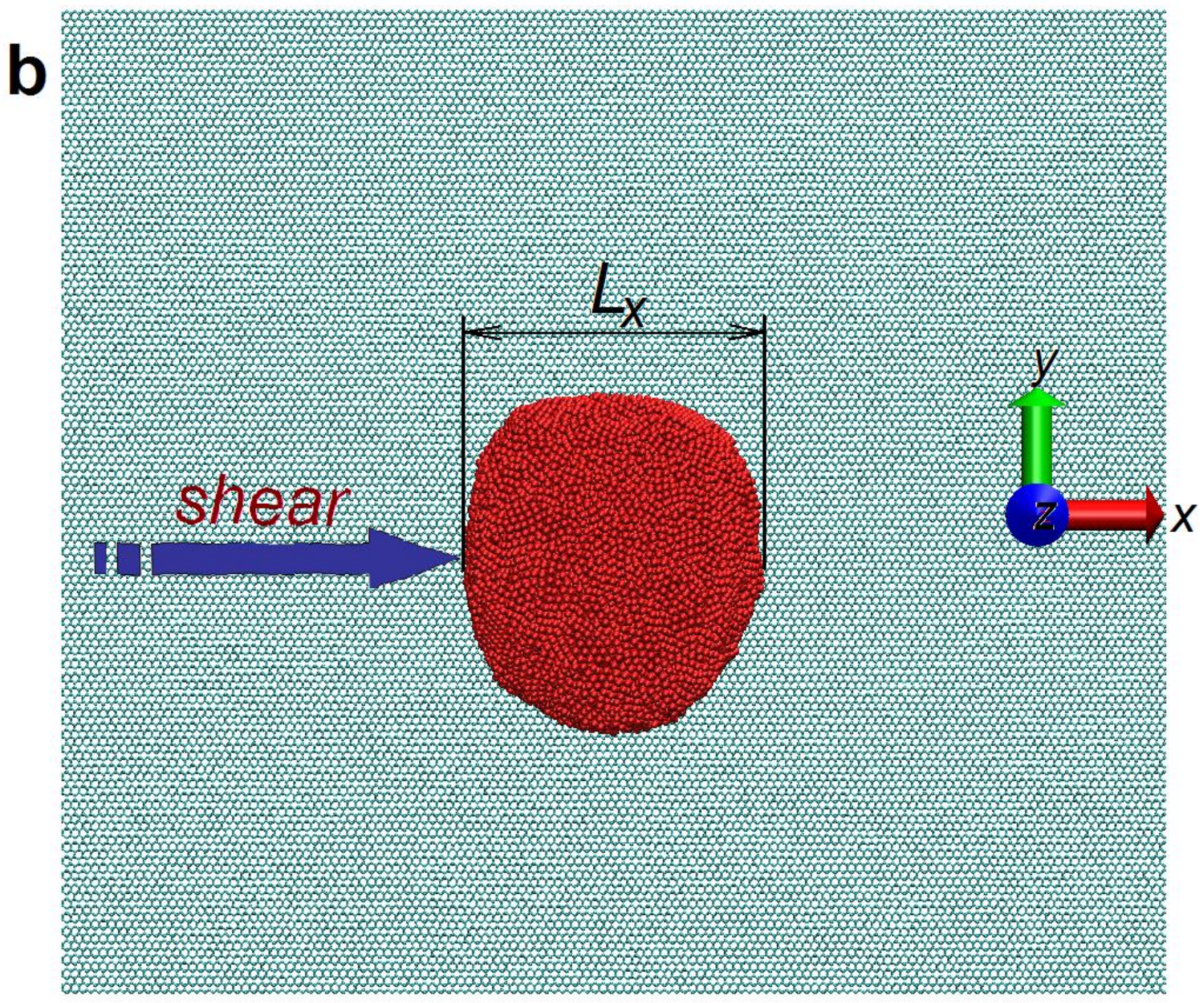}}
\caption{Snapshots of the formed nanoparticle containing 25,000 Ag atoms: perspective (\textit{a}) and top (\textit{b}) views.}
\label{fig1}
\end{figure}

Ideal vacuum conditions are maintained, so our simulations are not able to reveal the role of contaminations in friction of nanoparticles. We consider a graphene sheet lying in the \textit{xy} plane with zigzag and armchair edges parallel to \textit{x} and \textit{y} directions, respectively (see~\ref{fig1}, all snapshots in this work were produced with Visual Molecular Dynamics software~\cite{Humphrey1996}). To hold the sample in space, boundary carbon atoms along the perimeter of the graphene layer are held fixed throughout the simulations. Silver and nickel nanoislands containing from 5000 up to 30,000 atoms are considered. For each nanoparticle's size the unique \textit{x$\times$y} dimensions of the graphene sheet are used and they vary from about 19.68~nm$\times$17.04~nm to 36.40~nm$\times$31.52~nm, respectively. The total number of atoms involved in the calculations varies from 17,800 to 73,808.

Interactions between carbon atoms in graphene are described by the harmonic potential~\cite{Sasaki1996}.
%potential of the following form~\cite{Sasaki1996}:
%\begin{equation}
%\label{spring}
   % V_{C} = \frac{1}{2}\sum_{i-j}\mu_{r}(r_{ij}-r_{0})^{2}+
   % \frac{1}{2}\sum_{i-j-k}\mu_{\theta}r_{0}^{2}(\theta_{ijk}-\theta_{0})^{2}+
    %\nonumber\\
   % \frac{1}{2}\sum_{i-(j,k,l)}\mu_{p}
    %\left(\delta z_{i}-\frac{\delta z_{j}+\delta z_{k}+\delta z_{l}}{3}\right)^2.
%\end{equation}
%The description of the quantities in~\ref{spring} can be found %in~\cite{Sasaki1996}.
The simulations are not concerned with the formation or breaking of chemical bonds in the layer, so this potential form may be more appropriate for our task in comparison with more sophisticated potentials such as Brenner~\cite{Griebel2007,Khomenko2010,Prodanov2010} or ReaxFF~\cite{Sen2010} as it is less time-consuming. Forces between metal atoms are derived from the alloy form of the embedded atom method (EAM) potential~\cite{Zhou2001} which is well fitted to basic material properties. For the metal--carbon interaction the pairwise 6--12 Lennard--Jones (LJ) potential is employed. The values of the parameters $\varepsilon=0.8738\cdot10^{-2}$~eV, $\sigma=2.4945$~\AA~with the cutoff distance $r_{\mathrm{c}}=2.5\sigma=6.2363$~~\AA~from the work~\cite{Sasaki1996} are used for both metals. Such an inaccurate approach to the choice of values of the parameters is caused by the absence of the reliable data for the interaction energies of Ag and Ni atoms with graphene in literature. Although there exist a number of results obtained from the first-principles calculations~\cite{Shibuta2006,Smith2006,Matsuda2007} this data depends on the relative alignment of the atoms. But in the present problem since metal--C pairs conform in many different ways the use of mentioned data would not give the exact interaction energy. Moreover, ab initio techniques can overestimate values in several times~\cite{Smith2006}. Therefore, our approach is a good starting point which should provide a qualitatively correct but not quantitatively precise description of the system which should be relevant for metal nanoparticles that weekly bind with graphene~\cite{Lewis2000}.

The simulation code is implemented using NVIDIA$\circledR$ CUDA$^{\mathrm{TM}}$ platform~\cite{Meel2008,Anderson2008} which allowed us to carry out the computations on a single graphics processing unit (GPU) NVIDIA$\circledR$ GeForce$^{\mathrm{TM}}$ GTX 260. Algorithms for GPU based on the neighbor-list technique from Ref.~\cite{Anderson2008} with our own algorithm for binning atoms into cells are employed, the equations of motion are integrated using the leapfrog method~\cite{Griebel2007,Rapaport2004} with a time step $\Delta t=0.2$~fs.

\section{Results and discussion}
\label{results}

In the course of the simulations such quantities as temperature $T$ of the system, lateral position $X_{\mathrm{CM}}$ and velocity of the center of mass (CM) $V_{\mathrm{X}}$ of a nanoparticle, total shear force $F_{\mathrm{S}}$ are measured. Also frictional force $F_{\mathrm{f}}$ is defined as the sum of \textit{x}--components of forces acting on metal atoms from carbon ones. Dimensions $L_{\mathrm{X}}$ (cf.~\ref{fig1}), $L_{\mathrm{Y}}$, $L_{\mathrm{Z}}$ of a nanoparticle are computed as the difference between the coordinates of metal atoms with maximum and  minimum values along the corresponding direction. Structure of the nanoislands is characterized by radial distribution function~\cite{Rapaport2004} (RDF) $g(r)$ computed for time intervals of 1~fs. Although RDF is an integral characteristic and is not able to reveal the local ordering and subtle structural transitions in the medium, nevertheless it is valuable for general characteristic of structure of materials.

\begin{figure}[htb]
\includegraphics[width=0.51\textwidth]{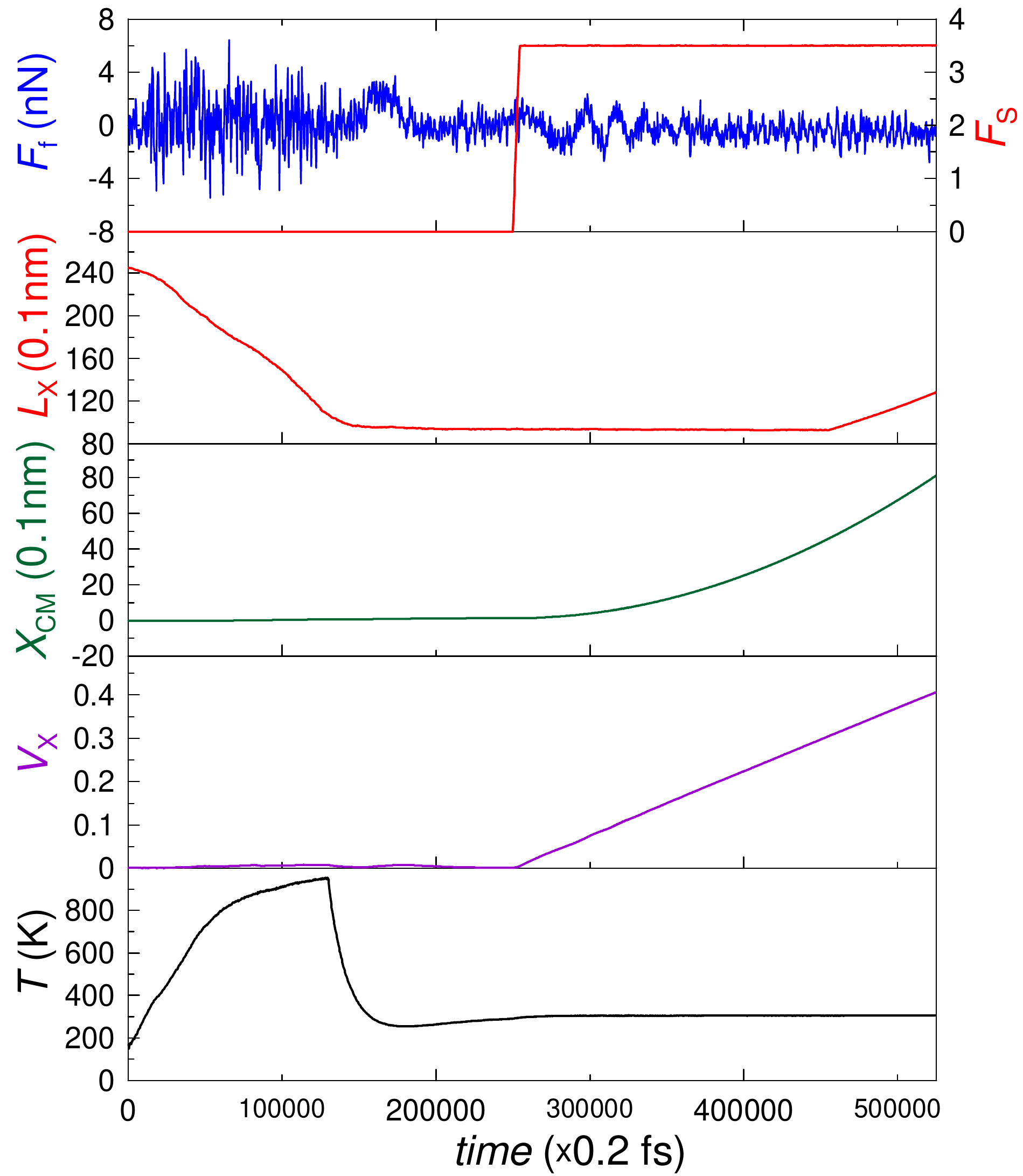}
\caption{Time dependencies of temperature $T$ of the system, lateral position $X_{\mathrm{CM}}$ and velocity $V_{\mathrm{X}}$ of the center of mass of the nanoparticle, total shear force $F_{\mathrm{S}}$, frictional force $F_{\mathrm{f}}$ and lateral dimension $L_{\mathrm{X}}$ obtained for Ni nanoisland containing 16,000 atoms.}
\label{fig2}
\end{figure}

In the experiments~\cite{Gnecco2007,Dietzel2008,Ritter2005} nanoparticles are obtained by vapor deposition. Our code allowed us to deposit metal atoms on a graphene sheet, but this approach is extremely time-consuming for the most of our system sizes. To avoid such a problem in this work the nanoparticles are obtained by the procedure imitating the dewetting of thin metallic films by thermal treatments~\cite{Geissler2010} and in more details it is as follows (movies Ni\_16000\_tot\_35136.avi, Ag\_25000\_tot\_63080.avi, Ag\_29000\_tot\_72800.avi are available). At the beginning of a simulation a slab consisting of several layers of metal atoms packed in the ideal FCC lattice is placed above the graphene layer. The vertical distance between the graphene plane and the lowest metal layer is 2.1~\AA~and 2.4~\AA~for Ni and Ag, respectively. Since many atoms in the slab are on its surface and have coordination which is smaller than the one in the bulk state, such ideal FCC lattice is not energetically profitable and metal atoms begin to rearrange into the more compact conformation corresponding to the minimal free energy. In~\ref{fig2} this is manifested in the decrease of the lateral size of the slab $L_{\mathrm{X}}$ with time. This process is accompanied by the release of the energy and therefore the temperature $T$ of the system is being raised (see the bottom plot in~\ref{fig2}) since at this stage the system is in microcanonical conditions, and the nanoisland melts. However, due to much smaller metal--C interaction energy comparatively with the metal--metal one, the contact angle of the forming metal cluster should approach 180$^\circ$~\cite{Shibuta2006}. The configuration with minimal energy will correspond to a ball, which is not suitable for our problem as it has small contact area. So, in order to obtain nanoparticles with the desired semispherical shape, at the appropriate time moment (in~\ref{fig2} this is 26~ps) defined empirically for each system size we begin to apply Berendsen thermostat~\cite{Griebel2007} both to metal and graphene during suitable time interval to cool the system down to the temperature of about 300~K. After that the thermostat is decoupled from the metal atoms and is applied only to graphene to dissipate the heat generated during the shear of the nanoparticle.

Procedure similar to the described above has been reported in Ref.~\cite{Neek2009} where the formation of metal nanoclusters on a graphene sheet for smaller system of 2500 metal atoms was investigated using classical MD. However, the obtained in Ref.~\cite{Neek2009} results showed the formation of a bundle of small clusters or even more or less continuous layers of atoms instead of one nanoparticle. Two main reasons may be argued for this. Firstly, very low constant temperature of 50~K maintained during the simulations might not allow dewetting of the metal atoms. Second reason may be the use of a pairwise interaction potential (instead of a many-body one) for metal atoms which is generally known not to be suitable for the description of metallic systems~\cite{Daw1984}.

\begin{figure}[htb]
\includegraphics[width=0.51\textwidth]{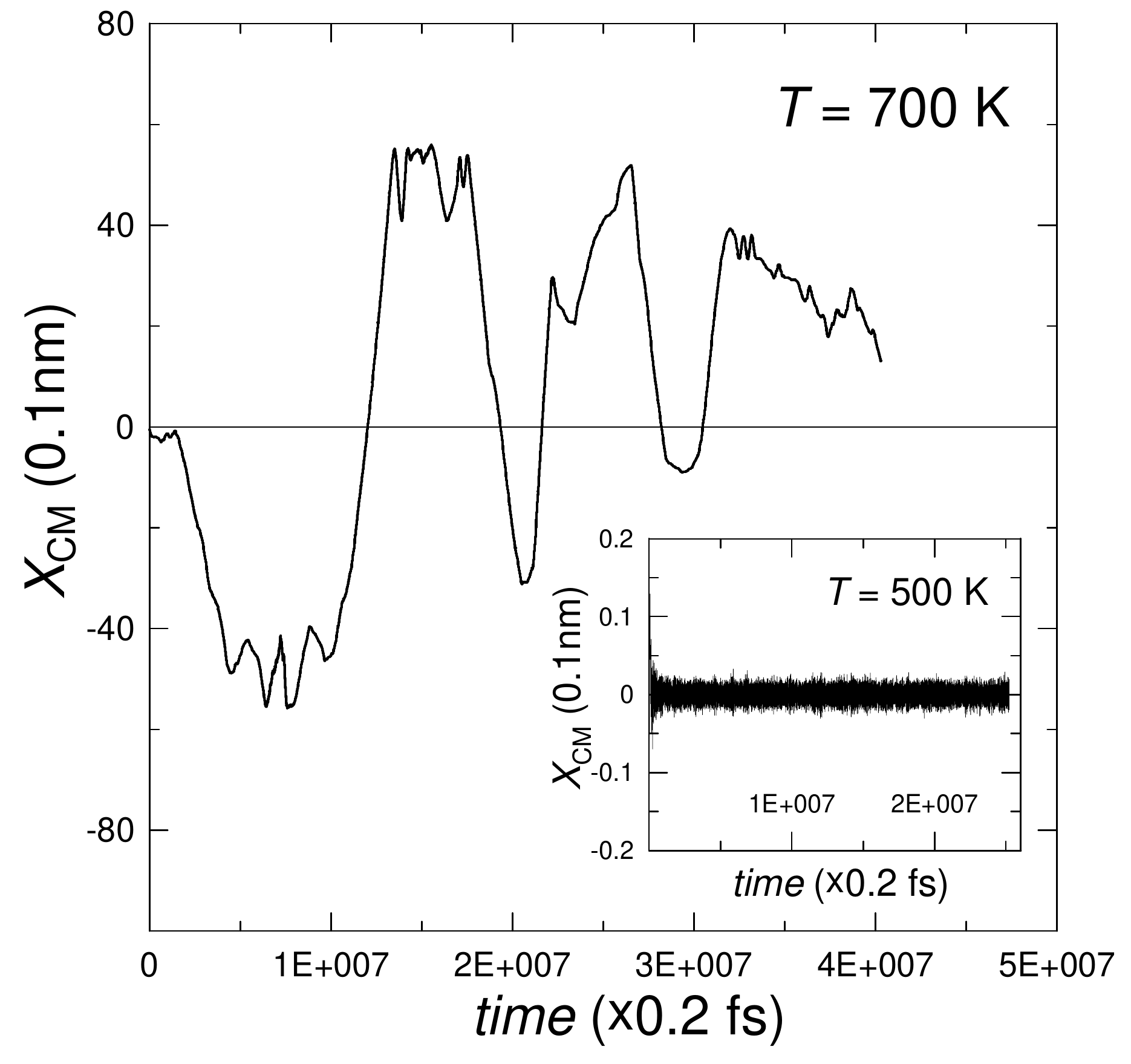}
\caption{Time dependencies of lateral position $X_{\mathrm{CM}}$ of the center of mass, obtained for Ni nanoisland containing 5000 atoms at 500~K and 700~K without shear.}
\label{fig3}
\end{figure}

In Refs.~\cite{Luedtke1999,Lewis2000,Yoon2003} it was found that small nanoparticles (with up to several hundreds atoms) are very mobile in the temperature range 300--900~K and diffuse easily on a graphite surface. Moreover, in Ref.~\cite{Lewis2000} the authors state, that assuming a proper scaling law for the dependence on size of the diffusivity, larger clusters containing up to 25,000 atoms should exhibit significant mobility. We decided to verify this, because high mobility of larger nanoclusters means that the energy of thermal fluctuations of atoms is comparable with adhesive energy and hence implies zero static friction. We considered a graphene layer with dimensions 15.74~nm$\times$13.63~nm containing 8192 carbon atoms interacting with Ni nanoisland consisting of 5000 atoms. The interaction potentials mentioned in previous section are employed, and it is important to note that in our case the magnitude of metal--C energy is about 2.5 times smaller than in Ref.~\cite{Lewis2000}. The behavior of the system at two values of temperature $T=500$~K and 700~K is explored for relatively long simulation runs. At 500~K the cooling stage spans the complete simulation time of 5~ns (or 25 million time steps). For the case of 700~K the cooling stage is 46~ps and the duration of the simulation is 8~ns (40 million time steps). These runs are not intended to comprehensively investigate the diffusion of the nanoparticle, but the aim is to ensure that in the conditions of our model the smallest nanoislands do not diffuse and therefore friction exists. As~\ref{fig3} shows, at $T=500$~K the value of $X_{\mathrm{CM}}$ has in fact zero value and is not changed significantly during the simulation, indicating the absence of diffusion. However, at 700~K the nanoparticle exhibits high mobility manifested in rather sharp changes of $X_{\mathrm{CM}}$ with time (movie Ni\_5000\_diffuse\_T700.avi is available). Thus, one may conclude that diffusion grows very rapidly with $T$. Nevertheless, in contrast to Ref.~\cite{Lewis2000}, our results suggest that it is unlikely for clusters containing more than 5000 atoms to be significantly mobile at $T$ less than 500~K. At about 300~K maintained during shear in our simulations significant adhesion and friction should exist.

After the cooling stage the shearing force is applied to the formed nanoparticle. In manipulation experiments with AFM pushing but not pulling is always involved~\cite{Resch1998,*Baur1998,Sitti2000,Dietzel2008}. To simulate pushing in our system shear force is applied along zigzag edge of graphene (which coincides with the \textit{x} direction) to all metal atoms with values of \textit{x}--coordinates that are smaller than the \textit{x}--coordinate of CM $X_{\mathrm{CM}}$ of the nanoparticle. At first, the force is incremented in steps of 0.0001~pN until the \textit{x}--component of the velocity of CM $V_{\mathrm{X}}$ reaches the value of 3.55~m/s. Then the shear force acting on each atom remains constant, and the simulations are held with constant total shear force $F_{\mathrm{S}}$. In most cases the nanoparticle is translated without rotation in the \textit{xy} plane.

\ref{fig2} partly already discussed shows typical time dependencies of the mentioned quantities for the Ni nanoparticle containing 16,000 atoms. It can be noted that after the beginning of shear $V_{\mathrm{X}}$ and $X_{\mathrm{CM}}$ increase linearly and quadratically in time indicating translation with constant acceleration corresponding to the constant $F_{\mathrm{S}}$. Frictional force $F_{\mathrm{f}}$ acting on the nanoparticle, however, is not uniform but has a sawtooth shape with growing in time frequency of spikes, which could indicate the stick--slip motion of the nanoparticle. Such a scenario may be the case for models where the objects are translated through the spring moved with constant velocity and where sawtooth force may lead to the intermittent behavior of $X_{\mathrm{CM}}$~\cite{Matsushita2005}. But in our case $X_{\mathrm{CM}}(t)$ has continuous parabolic shape due to constant shear force and therefore the nanoparticle moves continuously without intermittency.

\begin{figure}[htb]
\centerline{\includegraphics[width=0.51\textwidth]{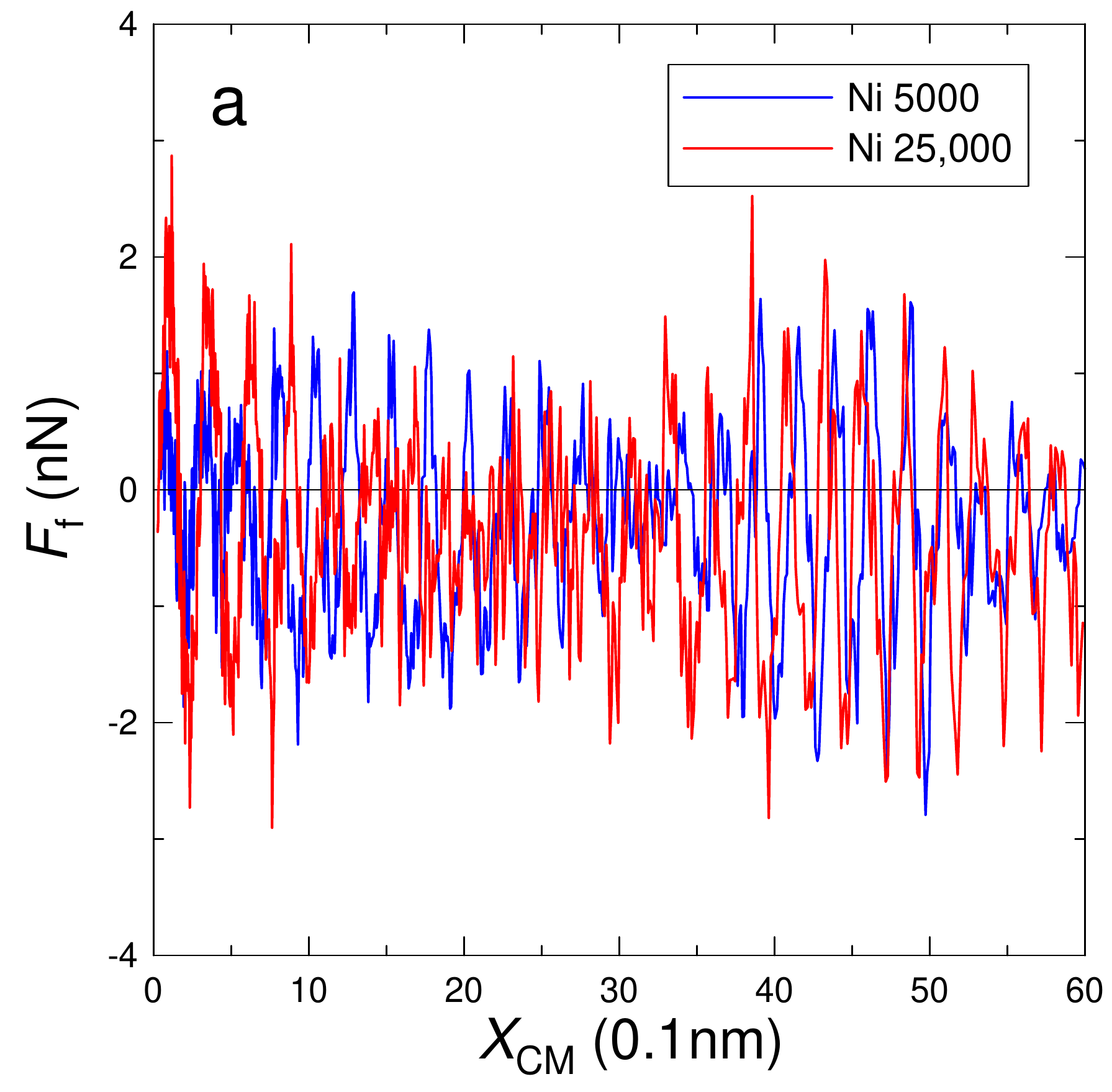}
\includegraphics[width=0.51\textwidth]{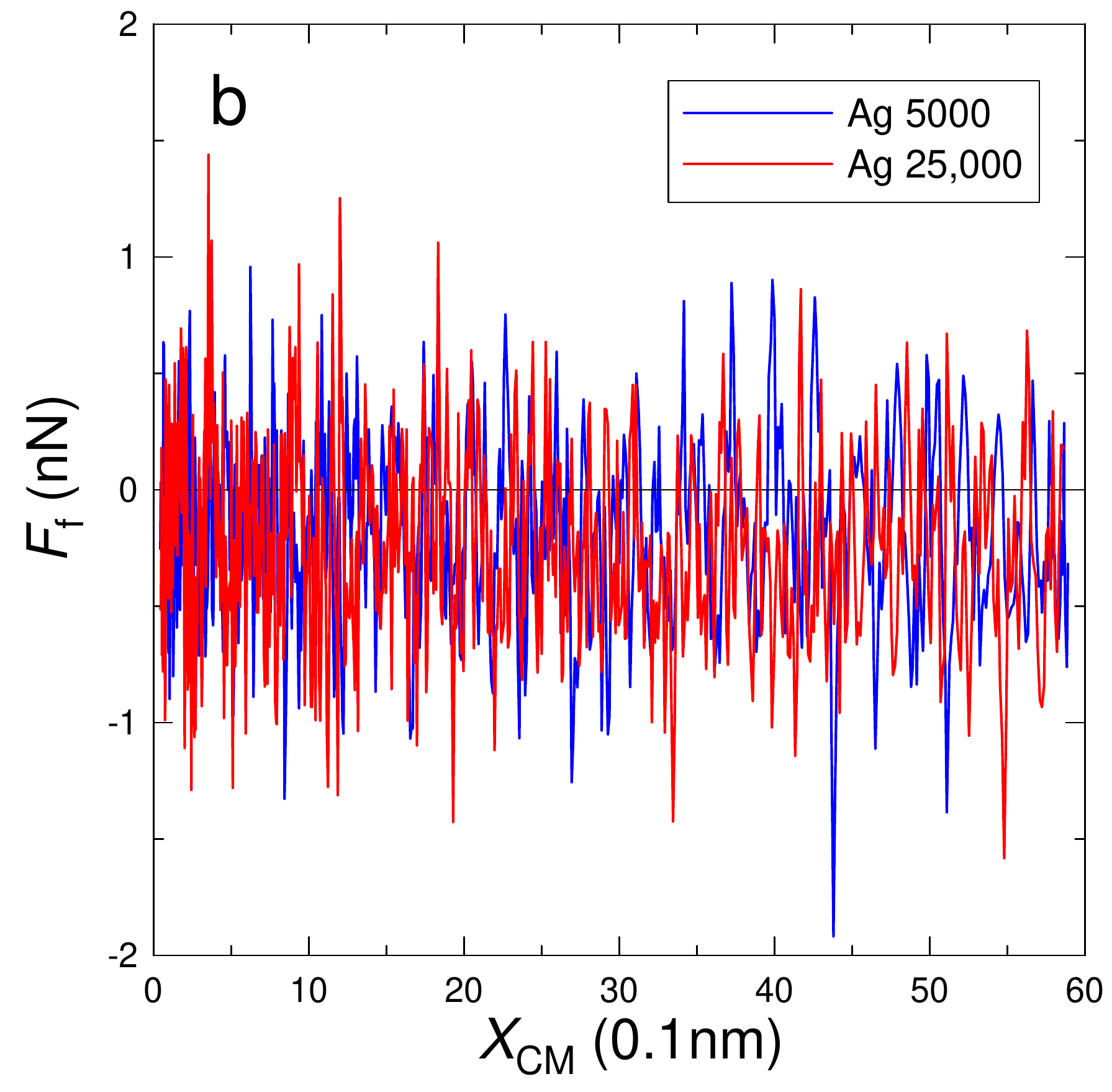}}
\centerline{\includegraphics[width=0.51\textwidth]{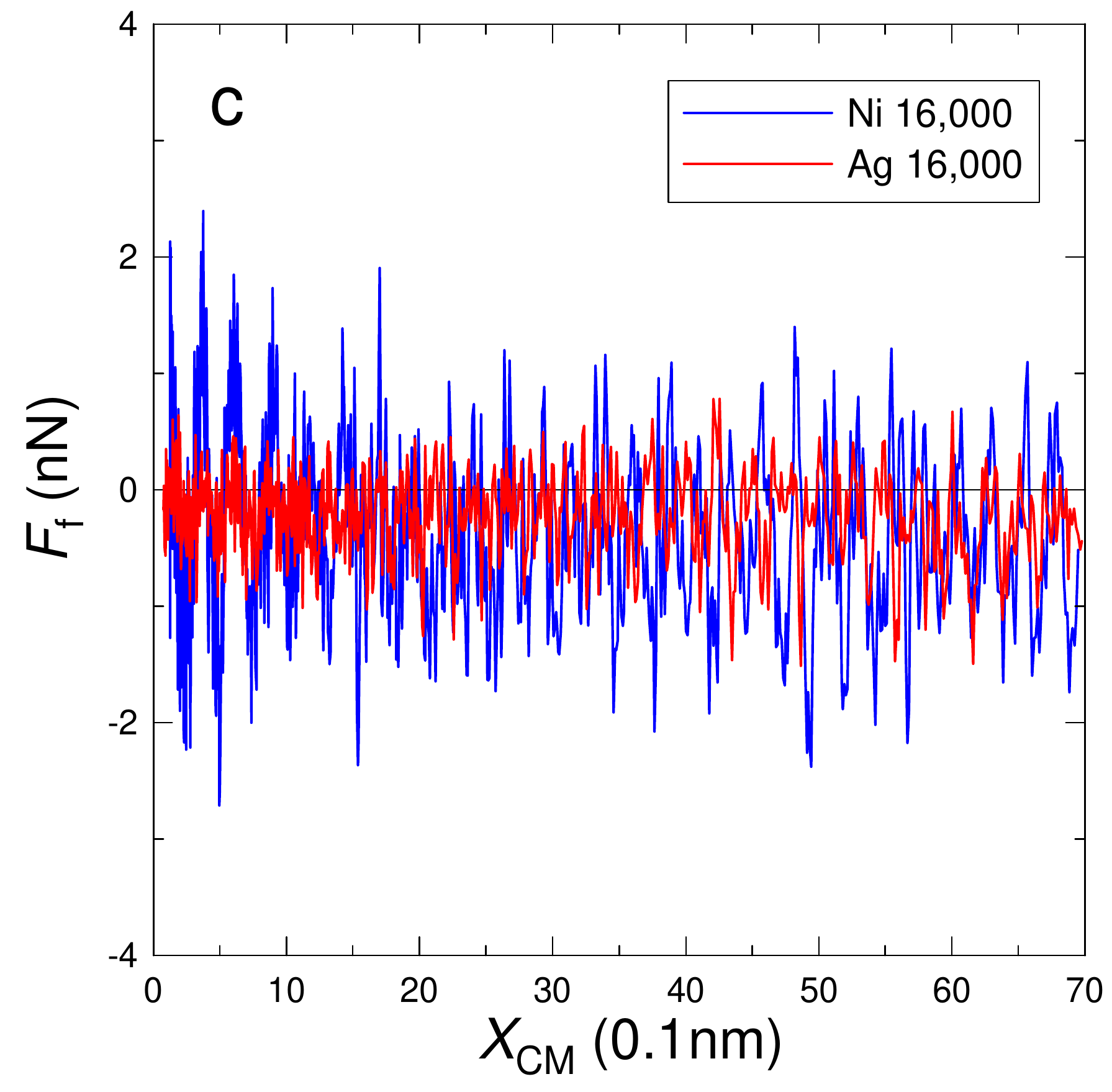}
\includegraphics[width=0.51\textwidth]{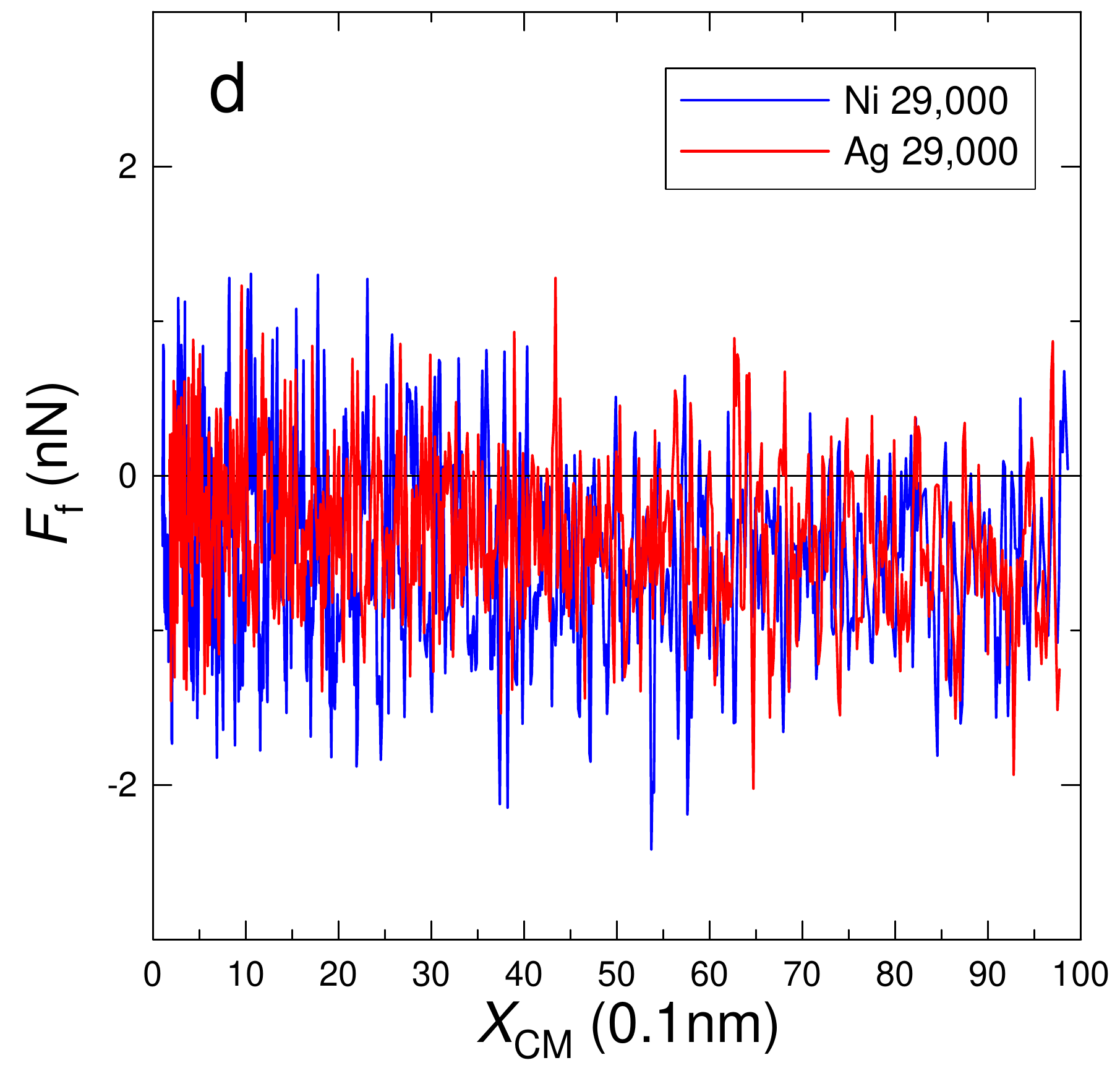}}
\caption{Friction force versus the lateral position of the center of mass of the nanoparticles: Ni with 5000 and 25,000 atoms (\textit{a}), Ag with 5000 and 25,000 atoms (\textit{b}), Ni and Ag with 16,000 atoms (\textit{c}), Ni and Ag with 29,000 atoms (\textit{d}). Only initial parts of plots are shown for clarity.}
\label{fig4}
\end{figure}

\begin{figure}[htb]
\centerline{\includegraphics[width=0.51\textwidth]{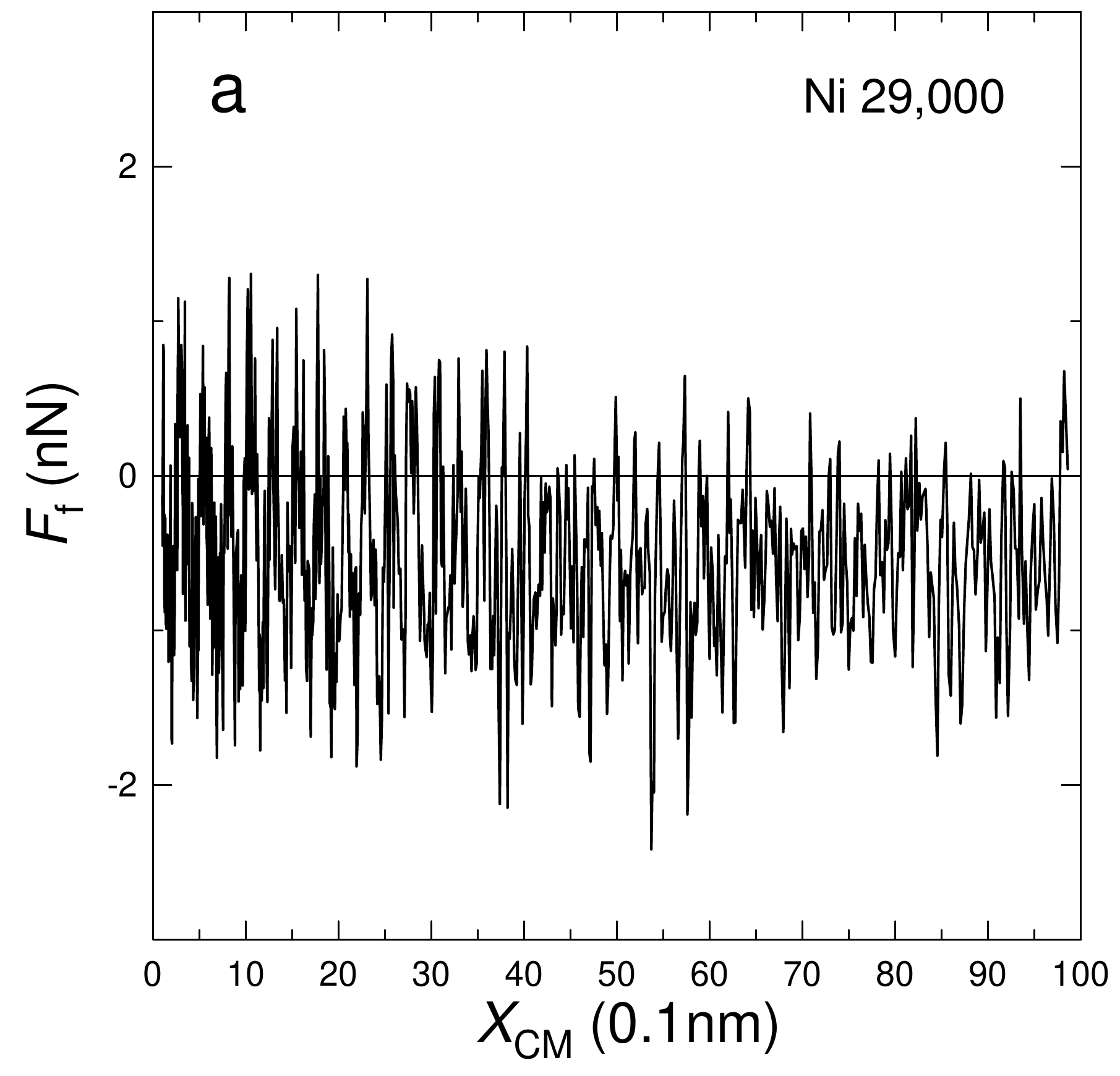}
\includegraphics[width=0.51\textwidth]{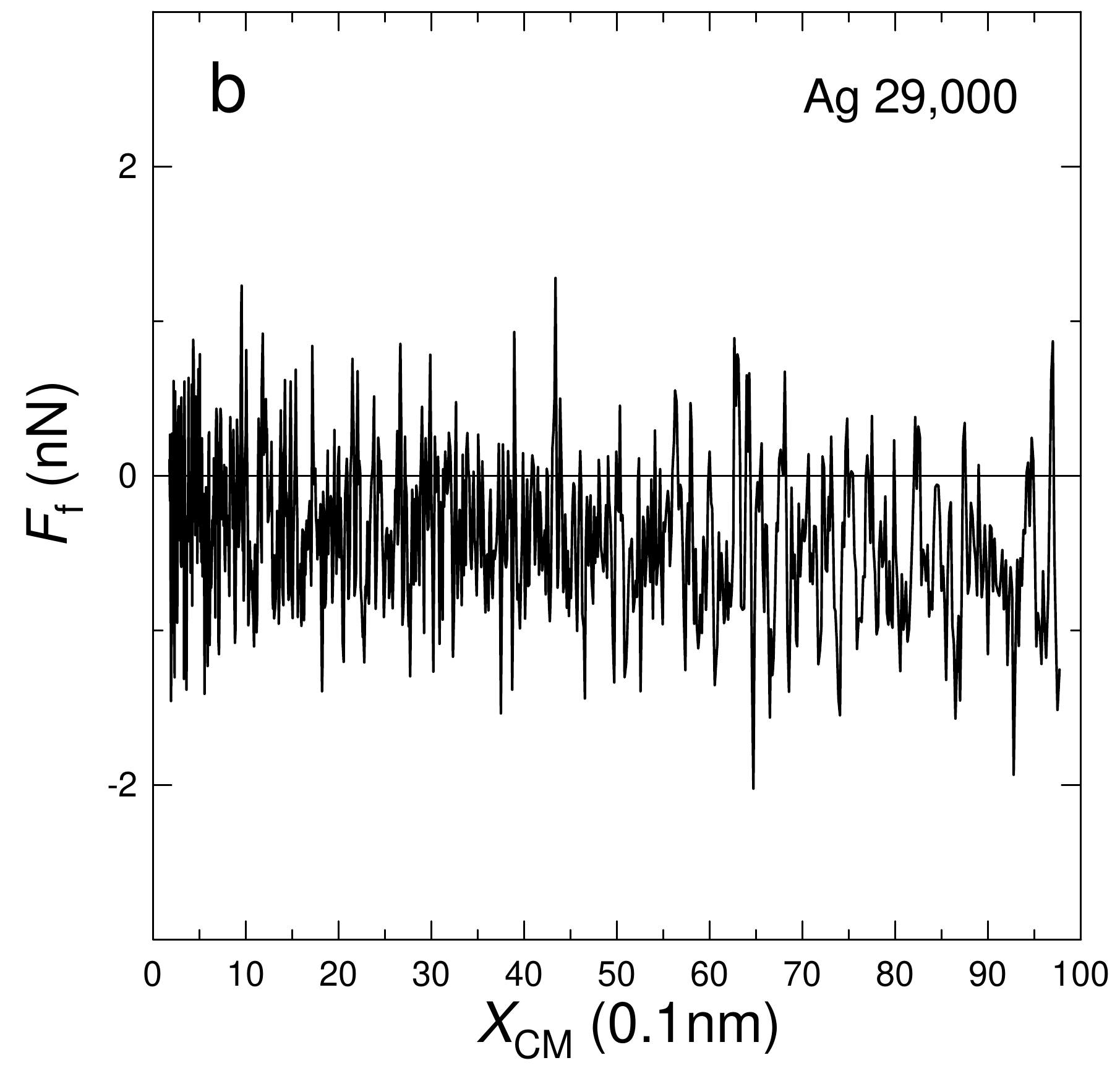}}
\caption{Friction force versus the lateral position of the center of mass of the Ni (\textit{a}) and Ag (\textit{b}) nanoparticles containing 29,000 atoms.}
\label{fig5}
\end{figure}

\begin{figure}[htb]
\includegraphics[width=0.51\textwidth]{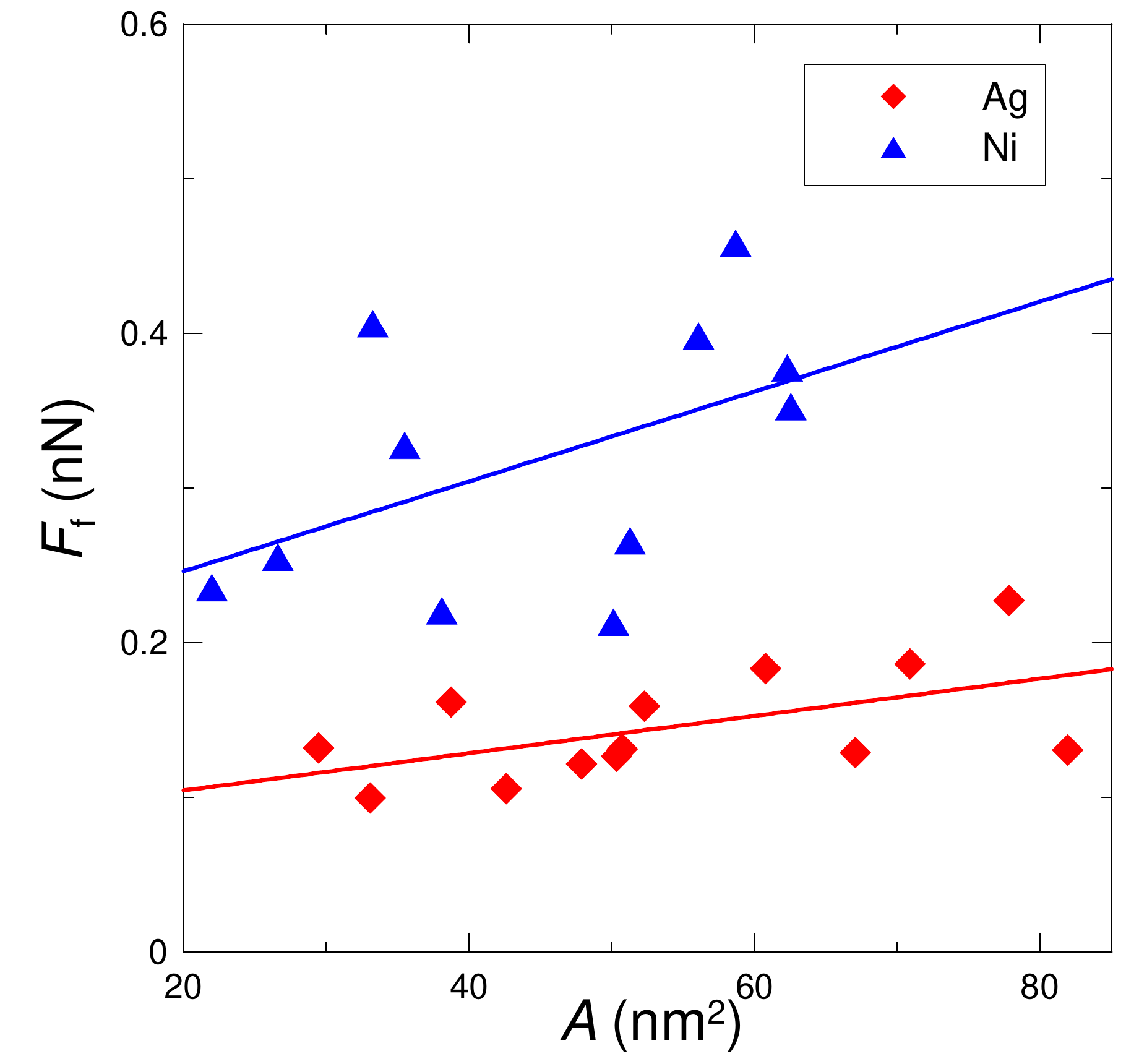}
\caption{Friction force versus contact area calculated for Ni and Ag nanoparticles.}
\label{fig6}
\end{figure}

To clarify the behavior of $F_{\mathrm{f}}$ let us consider \ref{fig4} and \ref{fig5} which plot dependencies of $F_{\mathrm{f}}$ on $X_{\mathrm{CM}}$ for several Ag and Ni nanoparticles. While for the most of Ni nanoislands friction force has a sawtooth form and spikes of $F_{\mathrm{f}}$ are more or less regular, for Ag friction force has irregular shape. The distance between spikes for Ni fluctuates near the value of the module of the lattice vector in graphene equal to 2.46~\AA~\cite{Castro2009} which may indicate the influence of the graphene lattice on the observed behavior. Note that positive values of $F_{\mathrm{f}}$ are observed suggesting that friction force changes direction during the nanoisland translation. On the average the amplitude of spikes for Ni is larger than for Ag. Time--averaged value of $F_{\mathrm{f}}$ (see~\ref{fig6}) grows approximately linearly with contact area $A$. For Ni relatively large data scattering is observed and $F_{\mathrm{f}}$ has lower value for Ag nanoparticles. Slopes of linear fits are 2.91~pN/nm$^2$ and 1.21~pN/nm$^2$ for Ni and Ag, respectively. The latter value is very close to the experimentally obtained 1.04~pN/nm$^2$ for Sb nanoparticles with areas more than about $10^4$~nm$^2$ sheared in ultrahigh vacuum~\cite{Dietzel2008}. Note, that in the experiments when Sb nanoparticles are translated on different surfaces (graphite and molybdenum disulfide) the obtained linear fits have the same slope. In our case for different metals but for the same surface the slopes are different.

The contact area $A$ in the simulations is defined similarly to the experiments using the lateral dimensions $L_{\mathrm{X}}$ and $L_{\mathrm{Y}}$ of the nanoparticle and approximating it with ellipse. Also note, that in \ref{fig2} the dimension $L_{\mathrm{X}}$ linearly grows in time when the nanoparticle approaches the boundary of the graphene layer (this was observed for all nanoparticles). To avoid the influence of this boundary effect on the results, averaging of $F_{\mathrm{f}}$ is performed from the moment when $F_{\mathrm{S}}$ is applied up to the beginning of this change of $L_{\mathrm{X}}$.

\begin{figure}[htb]
\includegraphics[width=0.51\textwidth]{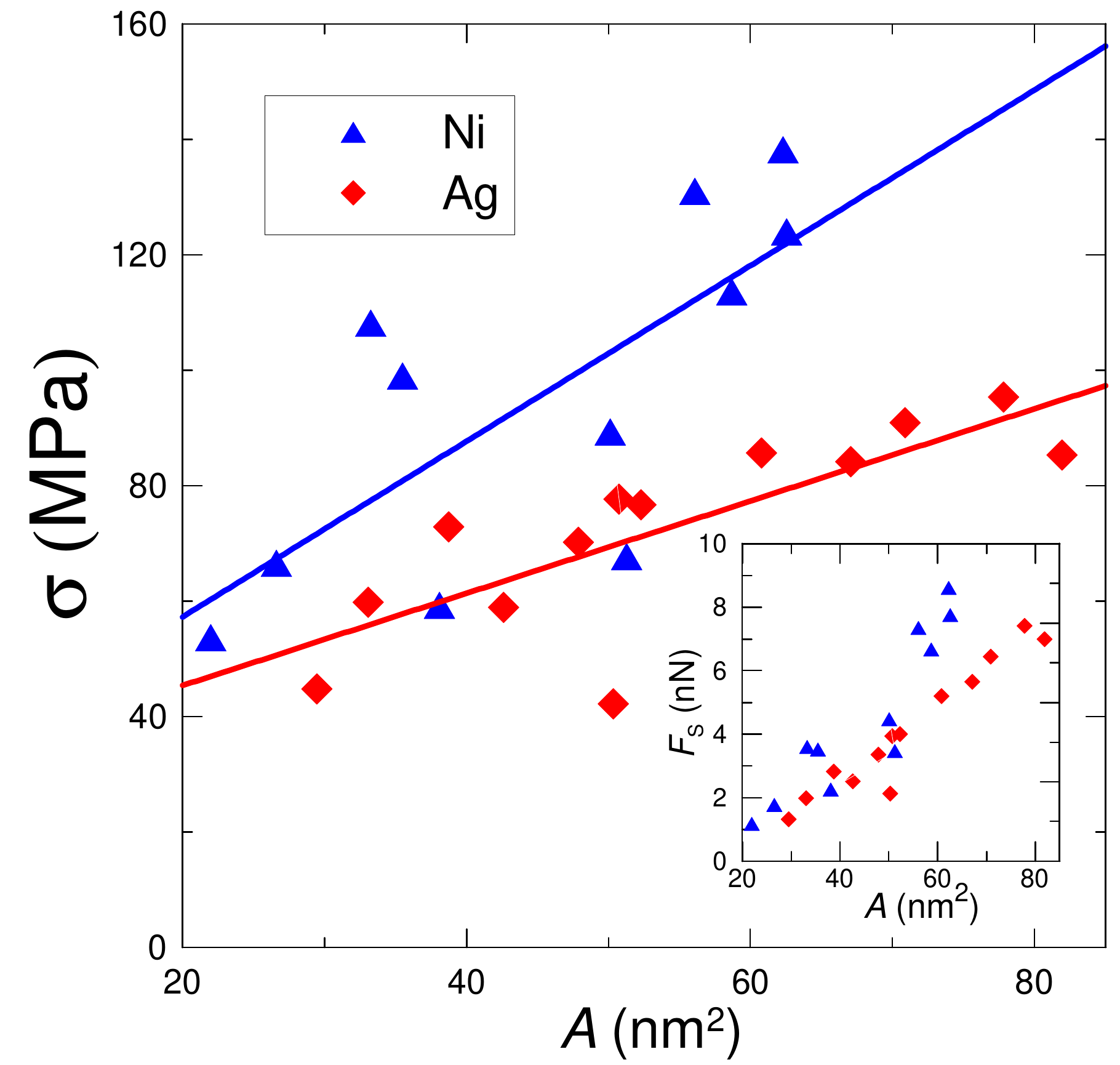}
\caption{Shear stress versus contact area calculated for Ni and Ag nanoparticles. Inset: shear force versus contact area.}
\label{fig7}
\end{figure}

Shear force $F_{\mathrm{S}}$ and shear stress $\sigma$ also increase approximately linearly with $A$ (cf.~\ref{fig7}). Values of $\sigma$ are on the order of $10-10^2$~MPa, which is closer to the experimental data obtained for large nanoparticles with areas more than about $10^4$~nm$^2$. Smaller nanoislands in the experiments as a rule require shear stresses on the order of 1~MPa and lower~\cite{Ritter2005,Dietzel2008}.

Analogous sawtooth shape of the lateral force is often observed experimentally for the tip of FFM (cf.~\cite{Socoliuc2004,Socoliuc2006} and references in~\cite{Khomenko2010}). In experiments~\cite{Socoliuc2004,Socoliuc2006} positive values of friction force acting on the FFM tip with radius of about 15~nm which is compared with the sizes of our nanoparticles were also reported. Additionally, although amplitudes of spikes in~\cite{Socoliuc2004,Socoliuc2006} are somewhat smaller than in our results, the average values of $F_{\mathrm{f}}$ are of the same order as obtained in the present simulations. Such a shape of $F_{\mathrm{f}}$ is quite good described by the Tomlinson--type models~\cite{Gnecco2007,Braun2006}. However, there are some discrepancies between the models and the experiments. One of the most significant is that these models are often based on the assumption that the tip contains a small number of atoms, ranging from one to several tens while in the experiments contact interface contains hundreds or thousands of atoms. The atoms are in general incoherently aligned with each other, and the periodic behavior should have been washed out since one would be averaging over many atoms out of phase with each other. The stick--slip motion with atomic periodicity should not be observed experimentally~\cite{Mate2002} but it takes place. This phenomena was detected in the first experiment using FFM with tungsten tip probing a graphite surface and at first it was suggested that the reason may be in the cleavage of a graphite flake~\cite{Khomenko2010,Mate2002}. But this explanation fell out of favor once researchers started observing atomically periodic friction on nonlayered materials where flake formation is impossible, and the question is still opened.

Our simulations show that sawtooth friction force can be expected during nanoparticle manipulation. However, it should be noted that the considered system differs from the FFM tip in the following aspect. It is known that accordingly to ``cobblestone model'' two factors contribute to friction on small scales: proportional to load and adhesive contribution proportional to the contact area~\cite{Khomenko2008}. While in experiments with FFM $F_{\mathrm{f}}$ can be considerably influenced by the external load~\cite{Socoliuc2004}, in our system this contribution is negligible. We estimated the gravitational force which could play the role of loading force to be about $5.3\cdot10^{-20}$~N for the heaviest Ag nanoparticle of 30,000 atoms. Hence this contribution can be safely neglected (as is done in the model) and therefore adhesive part completely prevails over the loading one.

The simulations also indicate that for the same metal--carbon interaction energy the presence of the sawtooth friction force and its average value for the given particle size may depend on the type of metal. Such a behavior may be caused by the collective atomic effects resulting from the combination of structural and elastic properties of the current metal--graphene system. Let us analyze possible factors such as structure and dimensions of the nanoparticles and elastic properties of graphene which may have impacted the obtained results.

\begin{figure}[htb]
\centerline{\includegraphics[width=0.51\textwidth]{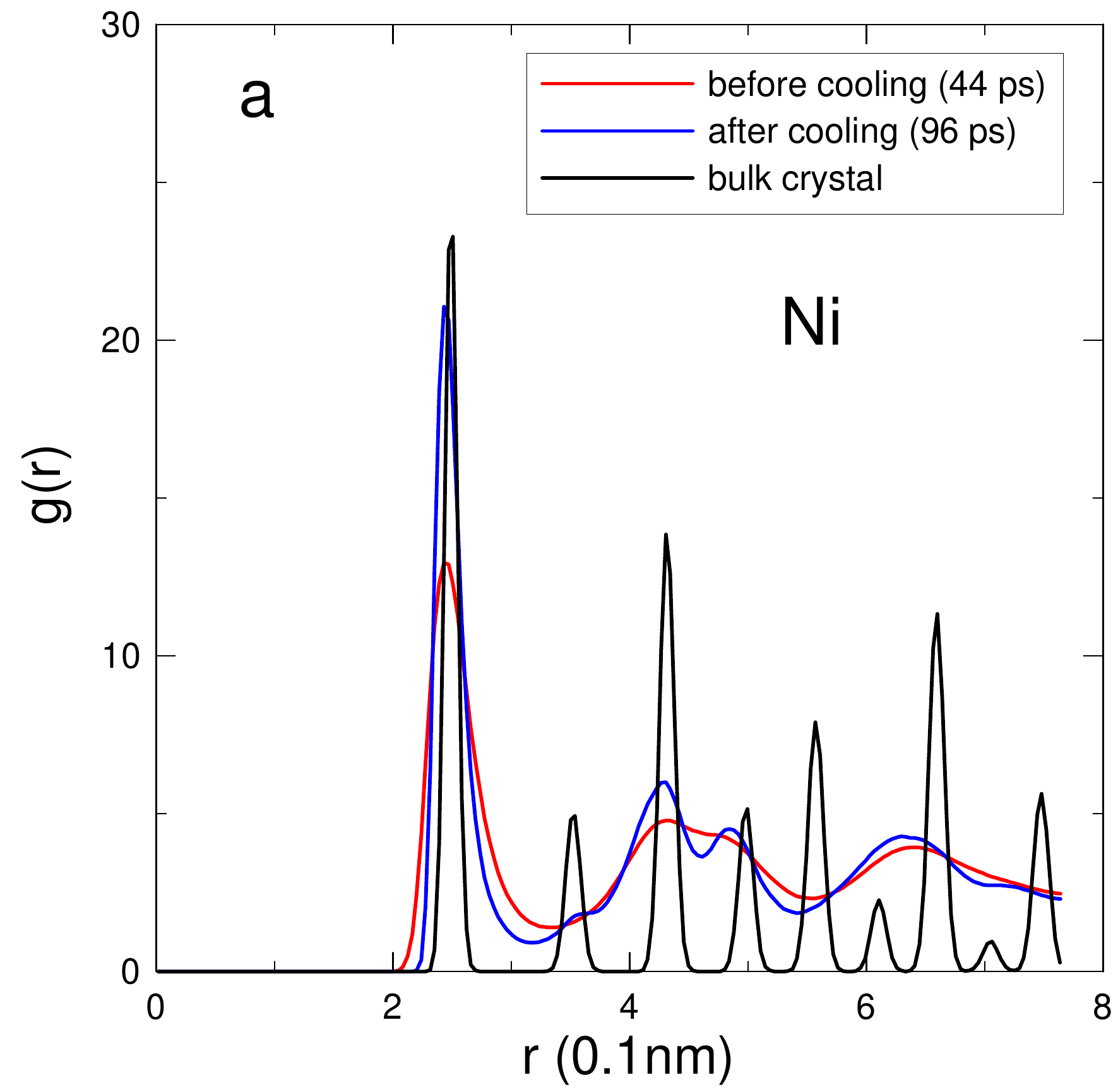}
\includegraphics[width=0.51\textwidth]{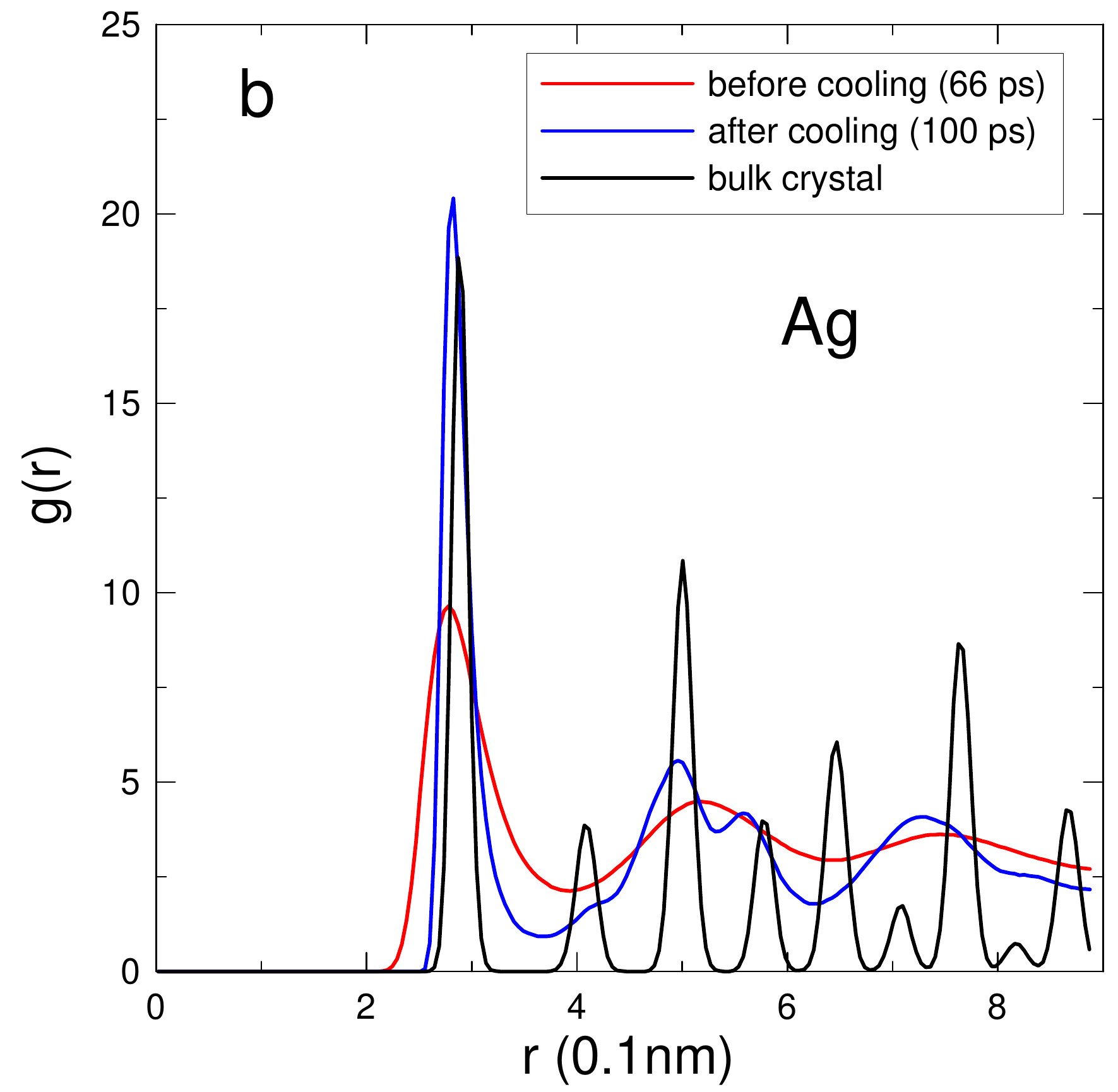}}
\caption{Radial distribution function obtained at different time moments for the Ni (\textit{a}) and Ag (\textit{b}) nanoparticles containing 29,000 atoms. Plots for bulk state are obtained using the same EAM potential.}
\label{fig8}
\end{figure}

\begin{figure}[htb]
\centerline{\includegraphics[width=0.45\textwidth]{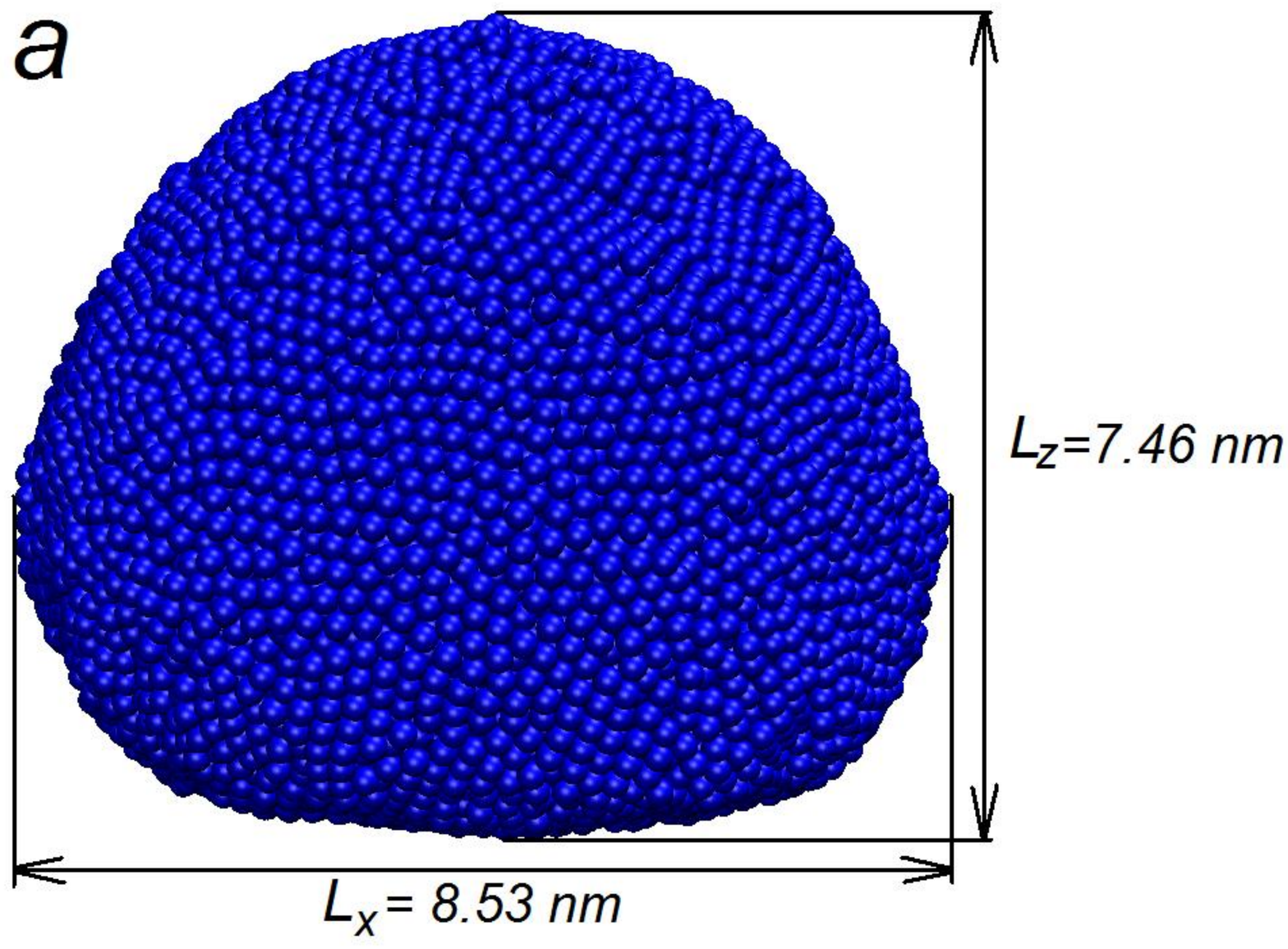}
\includegraphics[width=0.45\textwidth]{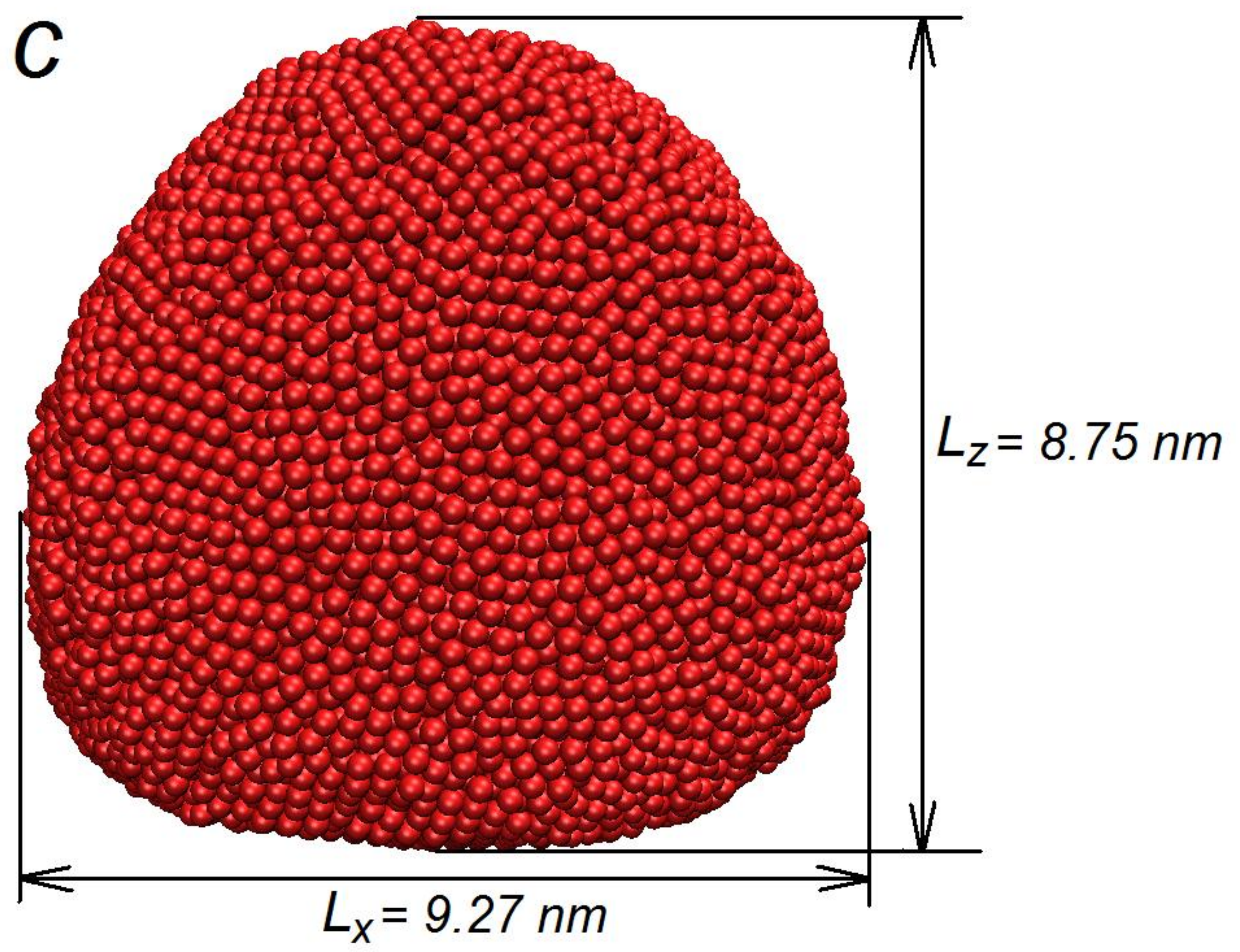}}
\centerline{\includegraphics[width=0.45\textwidth]{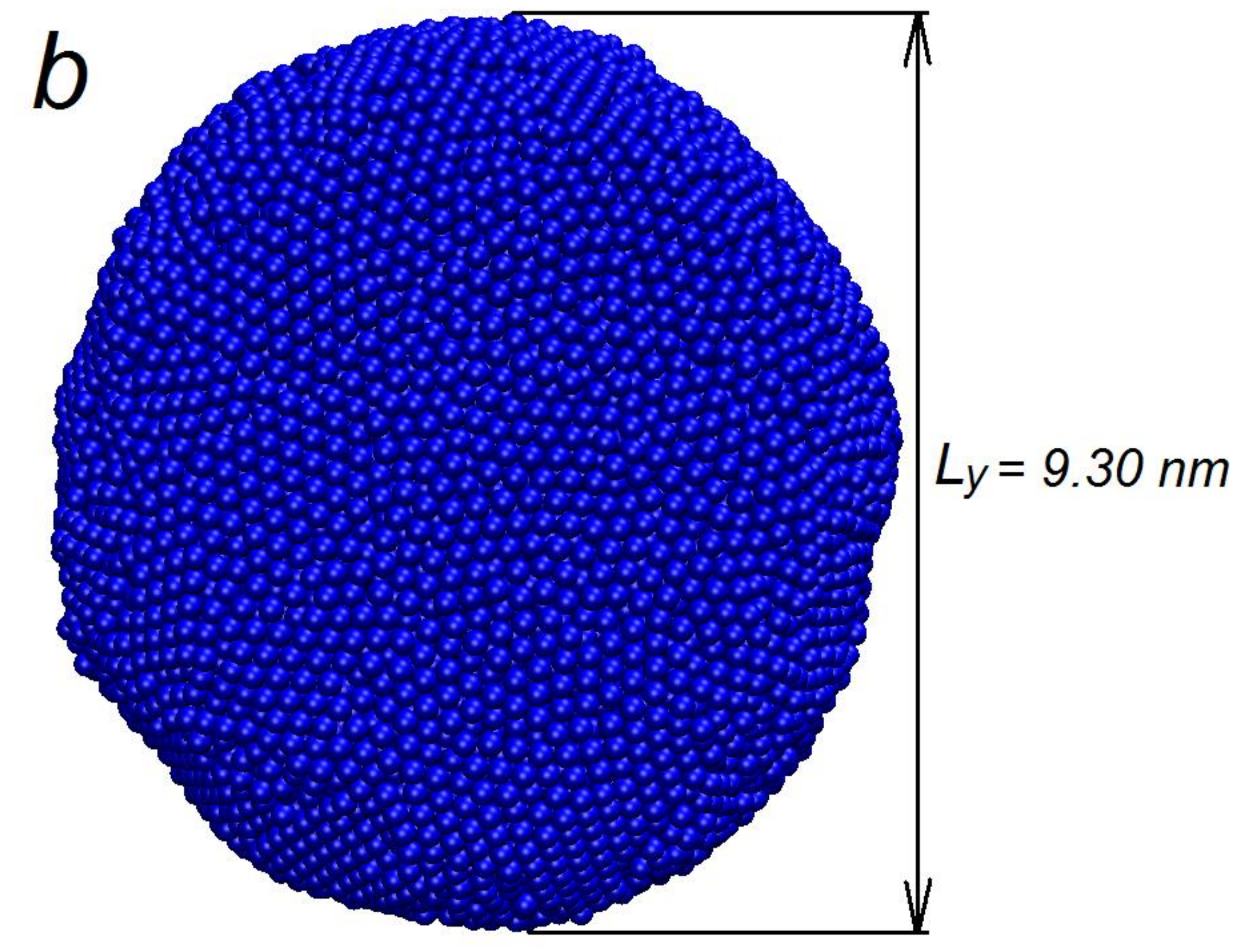}
\includegraphics[width=0.45\textwidth]{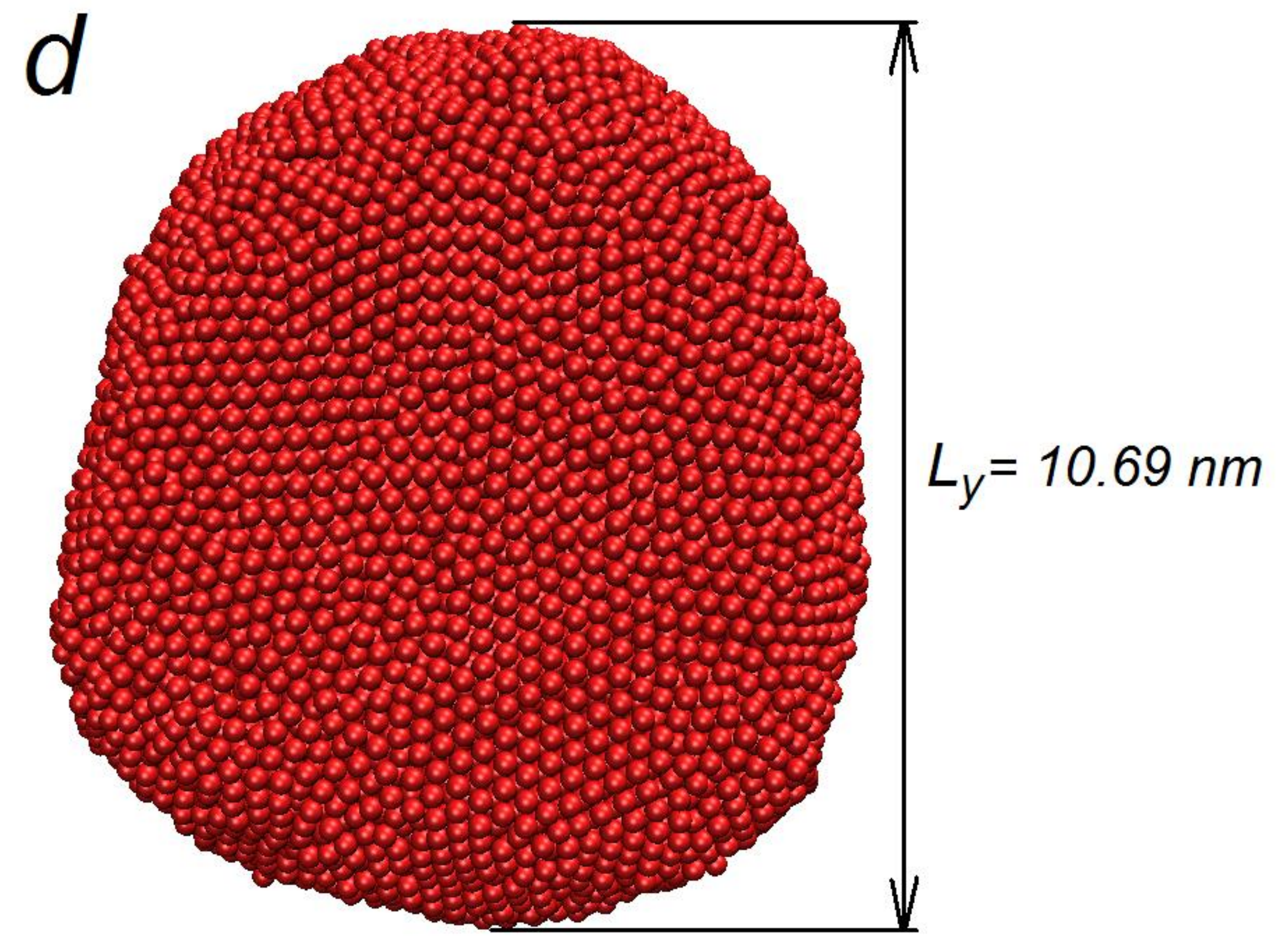}}
\caption{Side (\textit{a},\textit{c}) and bottom (\textit{b},\textit{d}) views of Ni (\textit{a},\textit{b}) and Ag (\textit{c},\textit{d}) nanoparticles containing 29,000 atoms.}
\label{fig9}
\end{figure}

In many analytical theories and seminumerical models structure of the surfaces is considered to be a crucial factor for the friction~\cite{Gnecco2007,Dietzel2008,Matsushita2005}. Sliding of ordered commensurate surfaces is predicted to occur via stick--slip motion~\cite{Matsushita2005} while for incommensurate disordered surfaces intermittent motion should not be observed. Analysis of the structure of the nanoparticles can be carried out using RDF measured at different time moments (see~\ref{fig8}) and snapshots of the contact surface of the nanoparticles in~\ref{fig9}. As can be seen from ~\ref{fig8}, after the formation of the nanoparticle RDF has completely smeared shape for both metals, indicating the disordered structure of the nanoislands. After the cooling phase some ordering is observed manifested in the formation of the higher first spike corresponding to the nearest neighbor distance in the bulk state of about 2.49~\AA~and 2.89~\AA~for Ni and Ag, respectively~\cite{MetalsHandbook}, and of additional smaller spikes. However, comparatively with ideal bulk crystal small spikes are very smeared which may indicate that nanoparticle is amorphous or has a polycrystalline structure. During shear the structure of the nanoparticle is not changed significantly (as follows from RDFs not shown here). Visual inspection of~\ref{fig9} shows that both Ni and Ag contact surfaces are disordered. Thus, we can conclude that generally the surfaces of graphene and of the nanoisland for both metals are incommensurate and the long--range order of the surfaces may not significantly influence the shape of the friction force. However, one can note that the average distance between the nearest neighbors in Ni near 2.49~\AA~is very close to the value of the lattice vector in graphene equal to 2.46~\AA~and to the distance between spikes in~\ref{fig4} and~\ref{fig5} in contrast to Ag that has larger distance. This may cause local commensurability and have an impact on the observed behavior.

In~\ref{fig9} one can see that the bottom part of the nanoparticles is deformed and due to wetting effects lateral dimensions do not exactly reflect the contact area. It is a well known fact that AFM images are convolutions of the samples and tip geometries, which gives a correct height but an overestimated width of the imaged features~\cite{Junno1995,Geissler2010} and this instrumental error should be taken into account. Although in the experiments for relatively large Sb nanoparticles~\cite{Ritter2005,Dietzel2008} authors state that the area obtained using AFM directly corresponds to the contact area and TEM inspection of the interface shows flat contact~\cite{Dietzel2008}, carefulness should be taken when extracting the contact area of the smaller nanoparticles from AFM images. Accordingly to the simulations, wetting effects due to small metal--surface interaction energy and the deformation of the substrate may cause the inconsistency between the measured by contour and the true contact area. This effect plus mentioned instrumental error may cause significant deviation of the measured results from the true ones.

Speaking about the dimension effect, as a rule adhesive interactions are assumed to be proportional to the area of contact~\cite{Khomenko2008}. However, in our simulations Ni nanoparticles that have smaller contact area than Ag experience larger friction. This may be attributed to the quantitative differences of the atomic structure of the materials leading to different surface energies of the nanoparticles. Smaller nearest--neighbor distance in Ni discussed above may cause the larger number of atoms to be located on the surface of the nanoparticle comparatively with Ag, and therefore may lead to higher surface energy and stronger adhesion. Additionally, metal atoms located upper than the bottom layer may give perceptible for such relatively small nanoparticles contribution to the adhesion as they lie in the range of the action of the LJ potential. Estimates of the interaction energy obtained from LJ potential at distance equal to two lattice constants of a metal give the contribution of about 6\% and 2\% of the energy minimum $\varepsilon$ for Ni and Ag, respectively. Thus, taking into account the disordered structure of the nanoparticle and the strain of graphene, it is probable that metal atoms located farther than the surface layer may give sustainable contribution into adhesion. And for Ni it should be several times larger than for Ag. Therefore, friction may depend not only on the area but also on the peculiarities of structure of the nanoparticle in the direction normal to the surface. This effect is beyond the capabilities of the standard seminumerical models, such as Tomlinson or Frenkel--Kontorova~\cite{Gnecco2007,Braun2006} where only one atomic layer is considered.

Lastly, a few words about the contribution of the dynamic behavior of the graphene layer to friction. Fixed atoms at the boundaries of the sheet caused the presence of waves in graphene reflecting from these atoms~\cite{Smith2006} and they were not completely damped by the thermostat. Although these waves exist both for Ag and Ni nanoparticles, smaller mass and structural peculiarities of the latter may be the reasons of higher sensitivity to the waves of the force acting on the Ni nanoparticles comparatively to Ag ones. This question should be elucidated in the future studies.

\section{Conclusions}
\label{conclus}

Atomistic approach employed in the present study allowed us to reveal nonuniform, in particular the sawtooth shape of the friction force acting on metal nanoparticles. The simulations indicate that for the maintained temperature of about 300~K there is no diffusion of the nanoislands with the considered dimensions. But the dependence of the diffusion on temperature may be very rapid and it should be investigated in future studies. The results also suggest that time--averaged friction force grows approximately linearly with area of contact for both metals, although for Ni large data scattering exists. The obtained slope of the linear fit for Ag of 1.21~pN/nm$^2$ is very close to the experimental value for large Sb nanoparticles, and calculated values of shear stress also correspond to the ones obtained in the experiment for larger nanoparticles. Generally, in spite of ultrahigh vacuum conditions, our simulations did not show the regimes with vanishing friction in contrast to experiments described in Ref.~\cite{Dietzel2008}.

It was also shown that the shape and the average value of the friction force may depend on the type of material of the nanoisland and these results are discussed in the context of the peculiarities of the structure of the nanoclusters and the dynamic behavior of graphene. The adhesion and hence the friction of small nanoparticles may depend not only on the contact area but also on the local structure of a material, in particular in the direction normal to the surface. Also our simulations suggest that for small nanoparticles the measured by AFM tip shape of the nanoisland might not be the reliable source for the definition of the contact area as there could be significant deviation due to surface strains and wetting effects. The waves on the surface may influence the friction and this question along with others are the subject of further investigations where the impact of the shear direction, elasticity of the layer, number of graphene layers, contaminations, interaction energy of metal--carbon atoms should be studied.

%% classes.  Rather than use \section, an appropriate macro is
%% provided that will always work.
%%%%%%%%%%%%%%%%%%%%%%%%%%%%%%%%%%%%%%%%%%%%%%%%%%%%%%%%%%%%%%%%%%%%%
\acknowledgement

This research was supported by the Fundamental Researches State Fund of Ukraine.

%% \suppinfo macro.
%%%%%%%%%%%%%%%%%%%%%%%%%%%%%%%%%%%%%%%%%%%%%%%%%%%%%%%%%%%%%%%%%%%%%
%\suppinfo

\bibliography{khomenko_prodanov_jpcc_2010}

\section{Figure captions}

Figure 1. Snapshots of the formed nanoparticle containing 25,000 Ag atoms: perspective (\textit{a}) and top (\textit{b}) views.

\vspace{0.3cm}
\noindent
Figure 2. Time dependencies of temperature $T$ of the system, lateral position $X_{\mathrm{CM}}$ and velocity $V_{\mathrm{X}}$ of the center of mass of the nanoparticle, total shear force $F_{\mathrm{S}}$, frictional force $F_{\mathrm{f}}$ and lateral dimension $L_{\mathrm{X}}$ obtained for Ni nanoisland containing 16,000 atoms.

\vspace{0.3cm}
\noindent
Figure 3. Time dependencies of lateral position $X_{\mathrm{CM}}$ of the center of mass, obtained for Ni nanoisland containing 5000 atoms at 500~K and 700~K without shear.

\vspace{0.3cm}
\noindent
Figure 4. Friction force versus the lateral position of the center of mass of the nanoparticles: Ni with 5000 and 25,000 atoms (\textit{a}), Ag with 5000 and 25,000 atoms (\textit{b}), Ni and Ag with 16,000 atoms (\textit{c}), Ni and Ag with 29,000 atoms (\textit{d}). Only initial parts of plots are shown for clarity.

\vspace{0.3cm}
\noindent
Figure 5. Friction force versus the lateral position of the center of mass of the Ni (\textit{a}) and Ag (\textit{b}) nanoparticles containing 29,000 atoms.

\vspace{0.3cm}
\noindent
Figure 6. Friction force versus contact area calculated for Ni and Ag nanoparticles.

\vspace{0.3cm}
\noindent
Figure 7. Shear stress versus contact area calculated for Ni and Ag nanoparticles. Inset: shear force versus contact area.

\vspace{0.3cm}
\noindent
Figure 8. Radial distribution function obtained at different time moments for the Ni (\textit{a}) and Ag (\textit{b}) nanoparticles containing 29,000 atoms. Plots for bulk state are obtained using the same EAM potential.

\vspace{0.3cm}
\noindent
Figure 9. Side (\textit{a},\textit{c}) and bottom (\textit{b},\textit{d}) views of Ni (\textit{a},\textit{b}) and Ag (\textit{c},\textit{d}) nanoparticles containing 29,000 atoms.

\section{List of figures}

\begin{figure}[htb]
\includegraphics[width=\textwidth]{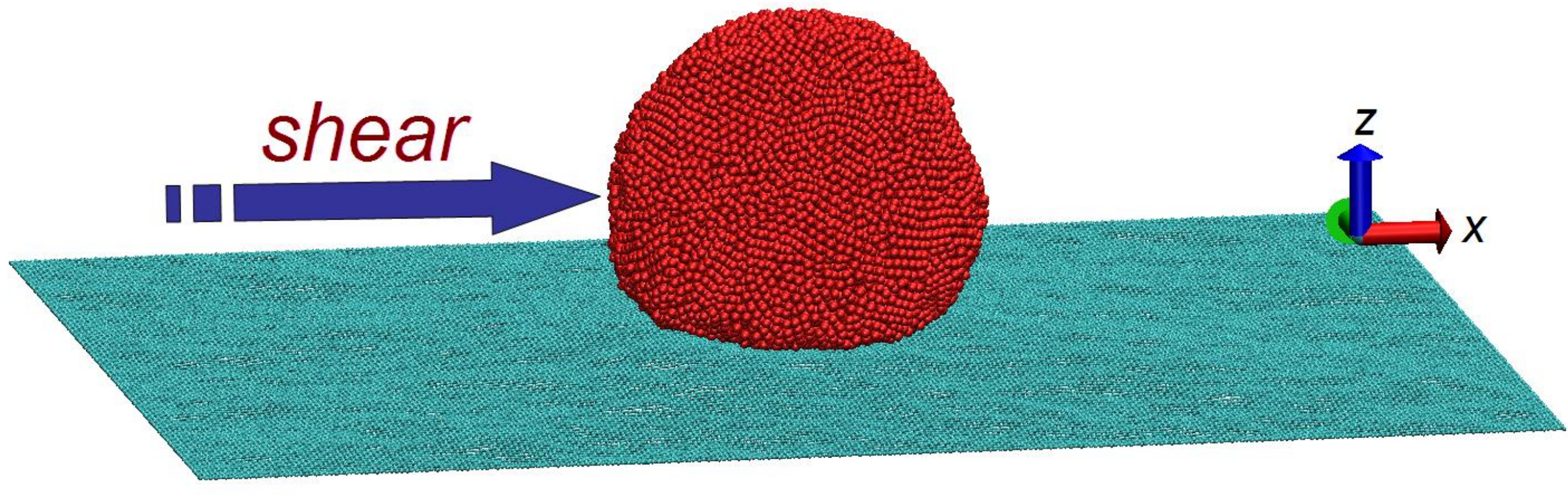}
\caption{Table of contents image.}
\label{TOC}
\end{figure}

\end{document}